# Toward an AI-Powered Computational Testbed for Workforce Policy

**Authors:** Sumer S. Vaid[1*] & Ashley V. Whillans[1]

**Affiliations:**
[1]Negotiation, Organizations and Markets Unit, Harvard Business School

*Corresponding author. Email: svaid@hbs.edu

**Abstract:** Workforce transformations are difficult to forecast and costly to mismanage. In particular, the integration of artificial intelligence into knowledge work currently affects a substantial share of the global workforce, yet this transition proceeds without tools to forecast how individual employees will respond psychologically and behaviorally. We combine recent advances in LLM-powered generative agents with foundational management science and organizational behavior research to propose *dynamic employee agents*. Among consenting populations, these agents can be seeded with HR records, validated psychometric measures, and digital activity data to simulate employees' cognitive, emotional, and behavioral trajectories across successive workdays during planned organizational changes. In this article, we detail the computational architecture required to construct this simulation platform and define the privacy, accuracy, and representativeness safeguards necessary for responsible deployment. We argue that establishing this prospective forecasting infrastructure is a critical technical requirement for managing the current global workforce realignment around AI.

Workforce transformations are notoriously difficult to forecast and immensely costly to mismanage. The integration of AI into knowledge work represents one of the largest such transformations in a generation (*1, 2*). Yet, no prospective methods exist to predict how individual employees will respond to these changes over time. Building on recent advances in AI (*3–5*) and on our experience developing and empirically validating such a platform, we introduce dynamic employee agents. We define dynamic employee agents as LLM-powered computational replicas (*3*, *4*) that capture how real-world employees think, feel, and behave. Unlike static (*6*) and domain-general computational replicas (*7*), dynamic employee agents are designed to evolve realistically over time and are built specifically to model workplace behaviors. A convergence of frontier model capabilities (*8*), accelerating enterprise AI adoption (*9*), and the absence of a reliable prospective forecasting infrastructure in organizations (*10*) makes this *the* critical window to build such a platform.

The need for such a platform is acute. Eighty-eight percent of organizations now use AI in at least one business function, yet only a third have scaled these programs beyond pilot stages (*9*). Employee mistrust in AI technologies continues to grow (*11*). New families of models and layered enterprise tools arrive faster than organizations can fully absorb them into their workforce (*12*). As a result, most enterprise AI pilots fail to deliver measurable impact (*13*), in part because few tools exist to anticipate employee responses before deployment.

**Moving Beyond Traditional Agent-Based Modelling**

Frontier large language models advance traditional agent-based modelling approaches (*14*) that rely on deterministic rules and pre-programmed behaviors (*15*). Trained on text spanning clinical case studies to nationally representative workplace surveys, large language models compress vast knowledge about human cognition, affect, and behavior into their parameters (*5*, *16*). When conditioned on human inputs such as interview transcripts (*7*), survey responses (*6*), or demographic profiles (*17*), large language models can reproduce individuals' self-reported attitudes and values (*6*), personality trait profiles (*18*), and choices in incentivized economic games (*19*). This capability has been demonstrated at increasing scale, from *in-silico* samples of a few dozen participants to large generative agent populations simulating tens of thousands of humans (*4*).

Our contribution here is twofold. First, we apply the generative agent architecture to the domain of workforce change, proposing a platform that predicts how individual employees within a specific organization respond to a proposed intervention before it is deployed. Second, we extend current generative agent approaches to explicitly model dynamics, producing computational replicas whose patterns of thinking, feeling, and behaving evolve across successive workdays in ways that track how real employees change. The integration of AI into knowledge work presents the most pressing potential use-case. Organizations redesigning knowledge work around AI can use such a platform to simulate the day-by-day trajectory of employee responses to AI adoption before any tool is deployed, from psychological states such as work engagement and team trust to behavioral outcomes such as tool use and collaboration patterns. Hence, the strongest use case lies in helping leaders evaluate the efficacy of AI tool roll-outs, a decision that currently rests on intuition (*20*), delayed pilot data (*13*), and retroactive cross-sectional surveys (*21*).

A dynamic employee agent platform can be constructed by drawing on three data sources (Figure 1, step 1). Human resource information systems can supply employees' role, team, and reporting relationships. Validated psychometric instruments can measure the employees' standing on constructs such as work engagement (*22*), team trust (*23*), psychological safety (*24*), and creativity (*25*). Macro-level instruments can capture global variables such as organizational climate and culture (*26*) and team norms (*27*). Together, these inputs determine who is simulated and at what resolution, with stakeholders selecting which data sources to draw on and at what cadence, in consultation with the workforce being simulated.

Each consenting employee can be represented as a computational replica built in two layers (Figure 1, step 2). A demographic prompt first conditions the base language model on population-level patterns associated with an employee's role, background, and context, drawing on the regularities documented in persona simulation research (*17*). Individualized psychometric data then constrain the replica to the employee's own measured standing on constructs such as work engagement and team trust, shifting predictions toward the individual employee.

Data collected by employee productivity tools (e.g., Microsoft Teams) can subsequently be routed into the simulated context of each agent (Figure 1, step 3, observed context). For instance, calendar data indicating that an employee spent the morning in back-to-back meetings could be ingested as a feature of the corresponding agent's simulated workday context, conditioning subsequent predicted behavior. Should this integration or data routing be undesirable or infeasible, LLMs can themselves be used to generate the organizational events each agent encounters during the simulation (Figure 1, step 3, synthesized context), including meetings, workload shifts, and interpersonal exchanges. Dynamic employee agents can also be nested within teams and social structures mimicking their real-world counterparts. For instance, agents in the same team might interact more frequently than agents on different teams.

In line with past research (*3*), the expectation is that such multi-agent simulations will exhibit emergent social behaviors that no individual agent was explicitly programmed to produce (*3*). Moreover, because the agents are powered by large language models, the simulated workforce can be queried by organizational leaders at will. Stakeholders can administer surveys on psychological constructs and conduct interviews on the reasoning behind observed behaviors, recovering the kinds of qualitative data that field research typically requires extended collection periods to obtain.

Proposed organizational changes can enter the platform as modifications to the "lived" experience of individual agents. For instance, an AI tool rollout can specify which agents gain access to which tools and when. The dynamic employee agent platform can hence translate a workforce change into a simulation protocol applied across the replicated workforce, producing predictions that can be compared across candidate interventions.

## Case Study: Testing AI Rollouts

In between-person experiments, each employee can be exposed to only one "treatment" at a time (*28*). A company weighing three AI rollout strategies must therefore choose one, implement it, and wait months for results without seeing how the other two would have fared on the same workforce. Field experiments handle the comparison by randomizing teams or offices to different conditions (*29*), but the logistical cost of these designs is extraordinary (*30*), and the firms willing to run them skew large and well-resourced (*31*), leaving the smaller workplaces where most employees work unstudied. The proposed dynamic employee agent platform helps leaders narrow the field of promising strategies before any rollout begins. A computationally replicated workforce can run every candidate in parallel, producing matched comparisons of how each employee's patterns of thinking, feeling, and behaving diverge across conditions.

The integration of AI into knowledge work fits this paradigm particularly well. Unlike interventions with predictable, localized or immediate effects, AI adoption is slow (*32*) and produces heterogeneous responses (*32*) across employees with different baseline attitudes and work patterns. These are precisely the conditions under which pre-deployment simulation adds value: effects are distributed across individuals and over time, candidate rollout strategies differ in ways whose consequences cannot be inferred from aggregate benchmarks, and the cost of choosing the wrong strategy compounds as deployment scales. The 95 percent enterprise pilot failure rate (*13*) reflects, in part, the mismatch between the diffuse workforce effects of AI and the narrow evidence base on which rollout decisions currently rest.

**Risks and Safeguards**

The capabilities that make generative agents a potentially powerful forecasting infrastructure also create new risks of privacy violations and algorithmic harm. We discuss three categories of protection against potential harms.

First, informed consent must be obtained from the employees whose digital traces and psychological data seed the simulation. Specifically, consent should be obtained at enrollment, refreshed whenever the simulation is repurposed for a substantially new question, and renewed with some level of cadence  reflects who the employee currently is rather than who they were years ago. The governing principle here is well established in adjacent domains (*33*), where data gathered for one stated purpose (for instance, improving AI adoption) cannot be redirected for other purposes (for instance, as a proxy for performance reviews) against the individuals who supplied it. System architecture, rather than policy language, should natively block simulation outputs from feeding performance reviews or disciplinary decisions. Critically, we argue that testing workforce changes on computational replicas of real employees occupies an ambiguous ethical position between pure modeling and behavioral experimentation on the employees themselves. The prudent default is to treat it as closer to the latter. Several practical safeguards follow from this. Organizations should prioritize keeping simulation outputs walled off from personnel records. Similarly, organizations should log queries made to the simulation platform, prohibit data merging by contract, and ensure that the system is open to third-party audit.

Second, regulators should require any simulation platform used to inform workforce decisions to be tested before deployment and disclosed publicly afterward. The vendor should demonstrate that the platform forecasts hold up against real-world data the model has never seen, that the forecasts cannot be easily manipulated or gamed, and that the methods and performance, including known failure modes, are available for independent inspection. The model for this is clinical trial registration, where the protocol and results enter the public record whether the findings are favorable or not.

Third, if an organization builds agents for a subset of the workforce and uses results to design organization-wide policy, those policies will favor the psychological profiles of employees who consented to participate. To the extent that consent and willingness to participate correlate with openness to experience, trust in management, or job security, resulting interventions could systematically misfire for the rest of the workforce. Developers of simulation platforms must demonstrate that results apply beyond the participants whose data seeded the agents, with explicit disclosure of coverage boundaries when the consenting employee population is not representative of the target employee population that will be subject to a workforce intervention.

Broader institutional support would strengthen these protections. For instance, organizations deploying AI systems that restructure work for large employee populations could file Behavioral Impact Assessments before implementation. These assessments can document simulation-based forecasts of psychological effects, disaggregated by role, team structure, and vulnerability profile, and disclosed to worker representatives. At the regulatory level, publicly maintained behavioral forecasting infrastructure, analogous to the economic forecasting capacity of the Congressional Budget Office, could allow policymakers to query predicted workforce effects of proposed labor regulations before enactment, from return-to-office mandates to right-to-disconnect legislation to large-scale reskilling programs. Where simulation-based evidence exists, employees affected by a planned restructuring should receive access to predicted psychological effects aggregated by role, team structure, and vulnerability profile before deployment begins, establishing a principle of informed exposure analogous to disclosure requirements in clinical research.

A further policy concern is the cultural generalizability of generative agents. Large language models display systematic bias toward simulating Western, educated, industrialized populations (*34*). For instance, LLM responses to psychological measures diverge from large-scale cross-cultural reference data, and model outputs become less accurate for populations outside Western industrialized samples (*17*). Techniques for reducing cultural bias span the end-to-end modelling pipeline, ranging from pre-training on multi-lingual corpora (*35*) to fine-tuning responses based on the preferences of diverse users (*35*). Building on decades of concerted effort to improve the external validity of psychological research, the behavioral sciences already possess large-scale publicly available datasets collected across multiple countries and organizational contexts. These resources can directly support the cultural alignment of generative employee agents. At the group level, protocols in which culturally diverse agents deliberate before producing a final response can help mitigate a collapse towards Western defaults (*36*).

**Validation and the Path Forward**

The value of dynamic employee agents is contingent on how well they predict real-world outcomes. Here, decades of psychological research can help specify validation frameworks and benchmarks that allow stakeholders to quantify the realism of observed simulations (*21*). For instance, a multi-faceted framework can assess whether the psychological states of simulated employees reproduce four psychometric descriptors of real employee data (*37*): the average levels of those states, how much they vary within a person from moment to moment, how different psychological states move together, and how gradually they evolve (and co-evolve) over time. Similarly, mapping the overlap between the behaviors of dynamic employee agents and their real-world counterparts can help identify the success and failure modes of the platform. Crucially, beyond simply emulating real-world patterns of change, validation must extend to causal benchmarks that test whether agents recover known treatment effects from historical quasi-experiments or documented organizational policy changes. Together, multilingual training, culturally grounded fine-tuning, organizational context, and individual-level seeding form a multi-pronged approach to improving cultural generalizability and increased realism of dynamic employee agents.

In conclusion, we critically note that dynamic employee agents earn their keep by lowering the odds of a failed pilot, not by replacing real-world pilots altogether. The use of dynamic employee agents should be treated as stage in the workforce transformation process, rather than a perfect substitute for human data collections *(38)*. Leaders can probe many candidate interventions in simulation, learn where these interventions falter and where they succeed, and subsequently carry sharper, better-evidenced versions into the field, where employees remain the final arbiters.

**Acknowledgments:** The authors would like to thank Yikun Chi for helpful discussion and feedback.

**Funding:** This work was supported by a Harvard Business School AI Institute Associates award to AVW and SSV.

**Author contributions:**
Conceptualization: SSV, AVW
Investigation: SSV
Writing – original draft: SSV
Writing – review & editing: SSV, AVW
Supervision: AVW

**Data, code, and materials availability:** Not applicable.

**Figure 1:** Building dynamic employee agents

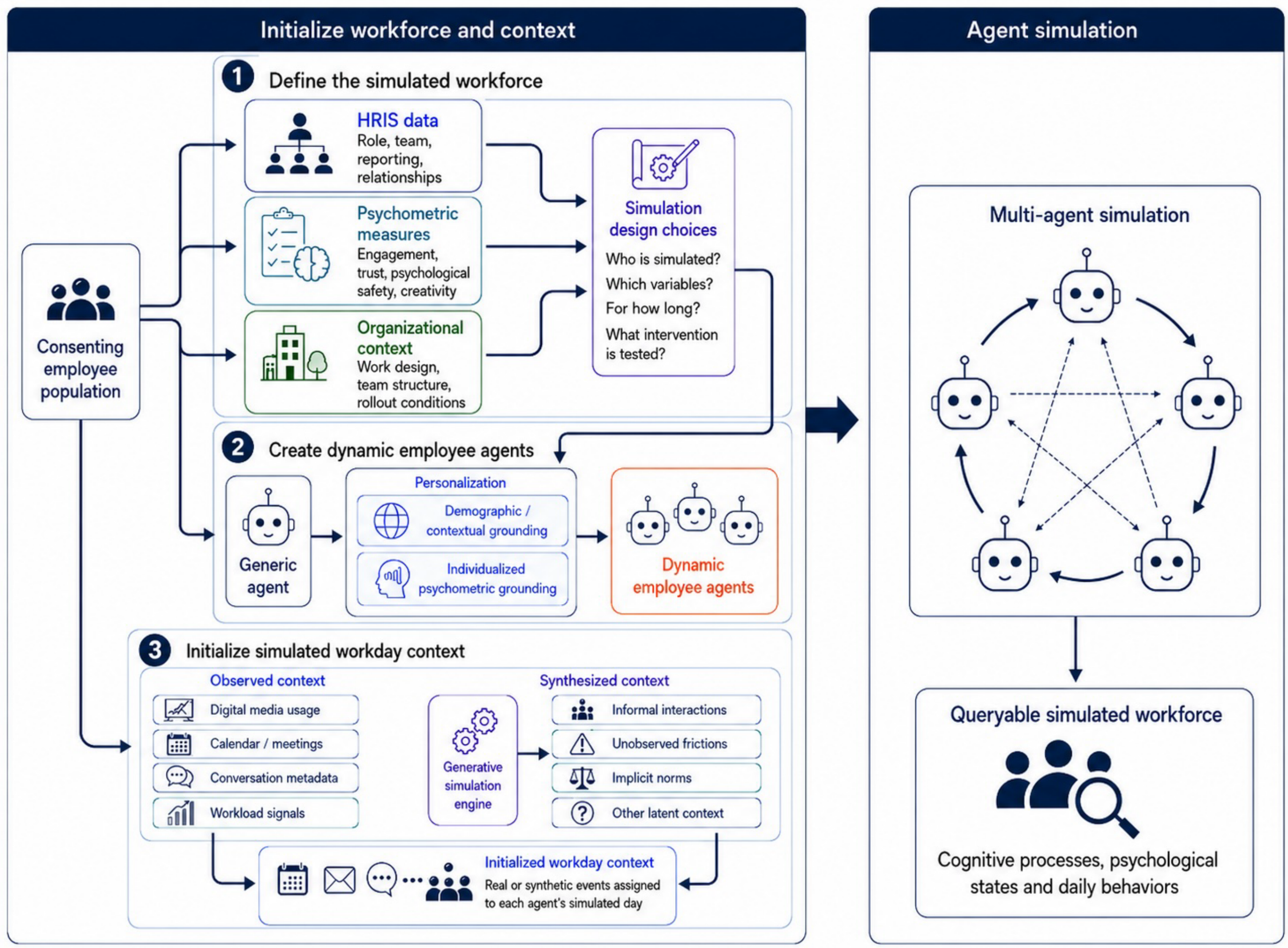